\documentclass{aa}
\usepackage{graphicx}
\usepackage{txfonts}

\usepackage{graphicx}
\usepackage{epsfig}
\usepackage{times}
\usepackage{natbib}
\usepackage{url}
\usepackage{color}
\usepackage{amssymb,amsmath}
\usepackage[%
  breaklinks = true,
  colorlinks = true,
  urlcolor   = blue,
  citecolor  = blue,
  linkcolor  = blue,
]{hyperref}
\graphicspath{ {./figures/} }
\begin{document} 
\title{A case--study of multi--temperature coronal jets for emerging flux MHD models}
\author{Reetika Joshi
\inst{1,2},
Ramesh Chandra
\inst{2},
Brigitte Schmieder
\inst{1,3,4},
Fernando Moreno-Insertis
\inst{5,6},
Guillaume Aulanier
\inst{1},
Daniel N\'obrega-Siverio
\inst{7,8},
\and
Pooja Devi
\inst{2}
}
\institute{Observatoire de Paris, LESIA, UMR8109 (CNRS), F-92195 Meudon Principal Cedex, France
\and
Department of Physics, DSB Campus, Kumaun University, Nainital -- 263 001, India\\
\email{reetikajoshi.ntl@gmail.com, reetika.joshi@obspm.fr}
\and
Centre for mathematical Plasma Astrophysics, Dept. of Mathematics, KU Leuven, 3001 Leuven, Belgium
 \and
University of Glasgow, Scotland
 \and
Instituto de Astrofisica de Canarias, Via Lactea, s/n, E-38205 La Laguna (Tenerife), Spain
 \and
Department of Astrophysics, Universidad de La Laguna, E-38200 La Laguna (Tenerife), Spain
 \and
 Rosseland Centre for Solar Physics, University of Oslo, PO Box 1029 Blindern, NO-0315 Oslo, Norway
\and
Institute of Theoretical Astrophysics, University of Oslo, PO Box 1029 Blindern, NO-0315 Oslo, Norway
}

 %
\abstract
{Hot coronal jets are a basic observed feature of the solar atmosphere whose physical origin is still being actively debated.} 
   {We study 
   six recurrent jets occurring in the active region NOAA 12644 
   on April 04, 2017. They are observed in all the  hot
   filters of AIA as well as cool surges
   in IRIS slit--jaw   high spatial and temporal resolution images.}
   %
  {The AIA filters allow us to study the temperature and the emission measure of the jets using the filter ratio method. We study  the pre--jet phases by analysing the  intensity oscillations at the base of the jets with the wavelet technique.}
   {A fine co--alignment of the AIA and IRIS data shows that the jets are initiated at the top of a canopy--like, double chambered structure with cool emission in one side and hot emission in the other. The hot jets are collimated in the hot temperature filters, have high velocities (around 250 km s$^{-1}$) and accompanied by the cool surges and ejected kernels  both moving at about 45 km s$^{-1}$.
   In the pre-phase of the jets, at their base
   we find quasi-periodic intensity oscillations in phase with small ejections; they have a period between 2 and 6 minutes, and are reminiscent of acoustic or MHD waves.}
    %
    {This series of jets and surges provides a good case--study to test the 2D and 3D magnetohydrodynamic (MHD) models that result from magnetic flux emergence. The double--chambered structure found in the observations corresponds to the cold and hot loop regions found in the models beneath the current sheet that contains the reconnection site. The cool surge with kernels is comparable with the cool  ejection and    plasmoids that naturally appears in the models.
   %
   }
 
   \authorrunning{Reetika Joshi et al.} 
   \titlerunning{Multi--temperature jets for emerging flux MHD models}
   \maketitle
%
\section{Introduction}
Solar coronal jets are detected along the whole solar cycle in a large wavelength range, from X-rays
\citep{shibata1992}  to the EUV  \citep{wang1998,alexander1999,innes2011,sterling2015,chandra2015,BJoshi2017}. 
Many are seen as collimated plasma 
material flowing along open magnetic field lines with high velocity. 
 Other interesting ejections are
 cool surges, which 
 emerge 
in the form of unwrinkled threads of dark material in H$\alpha$ \citep{Roy1973,Mandrini2002,Uddin2012,Li2016} and sprays, which are very fast ejections having their origin in 
filaments generally in active regions \citep{Warwick1957,Tandberg1980,Pike2002,
Martin2015}. In fact, some surges are closely related to hot jets \citep{Schmieder1988,canfield96}.
Solar coronal jets are observed in active  regions \citep{Sterling2016,Chandra2017,Joshi2018} as well as in quiet regions \citep{Hong2011, Panesar2016}.
Their physical parameters such as height (1--50 x 10$^{4}$ km), 
lifetime (tens of minutes to one hour),
width (1--10 x 10$^{4}$ km), and velocity (100--500 km s$^{-1}$) have been studied by many authors
\citep{Shimojo1996,Savcheva2007,Nistico2009,Filippov2009,Joshi2017}.

Magnetic reconnection is believed to be  the triggering mechanism
 behind the activation of the jet phenomenon according to different theoretical models  \citep{Yokoyama1995, Archontis2004, Archontis2005, Pariat2015}. 
Reconnection is a process of restructuring of the magnetic field lines and 
can occur in 2D 
\citep{Filippov1999, Pontin2005} or in 3D configurations
\citep{Demoulin1993,Filippov1999,Longcope2003,Priest2009,Masson2009}.
 In a 2D magnetic null point configuration, magnetic field lines contained in a plane and with opposite orientations come toward each other across an X--point 
 and change connectivity instantaneously; the result are hybrid field lines that are expelled away from the X--point, typically with velocities of order the Alfv\'en speed.
In 3D there is a whole variety of possible patterns (like: spine-fan, torsional, separator reconnection, etc); in many cases the underlying structure is what is known as a fan-spine configuration around a central null point. The field lines from inside the fan surface are joined to open field lines from just outside with ensuing connectivity change. Changes in the remote connectivity of magnetic field lines may also take place in regions with strong spatial gradients of the field components called quasi-separatrix layers (QSLs) \citep{Mandrini2002}. 
\\
Magnetic reconnection can take place as a result of a process of magnetic flux  emergence from the low solar atmosphere or interior. 
In typical magnetic flux emergence processes, 
the emerging magnetized plasma interacts with the pre--existing ambient coronal magnetic field, thus providing a favorable condition for magnetic reconnection, and therefore, for the occurrence of solar jets.
The observations indicate that the expansion of the magnetic flux emerging region leads to reconnection with the ambient quasi potential field and magnetic cancellation 
\citep{Gu1994,Schmieder1996,Liu2011, Guo2013}.
A number of numerical models have simulated this process
(see, for example \citealt{Yokoyama1996, Archontis2004, Moreno2008, Torok2009, Moreno2013, Archontis2013, Nobrega2016, Ni2017}).
In the model by \citet{Moreno2008}, in particular, a split-vault structure is clearly shown to form below the jet containing two chambers: the chamber containing previously emerged loops with a decrease in volume and the chamber containing reconnected loops with a increase in volume due to reconnection. This structure is also confirmed in radiation--MHD simulations by \citet{Nobrega2016}.
 The observations that motivated those models were either X-ray jets observed by Hinode \citep{Moreno2008},  
 or cool surges observed in chromospheric lines and bright bursts in transition region lines \citep{Nobrega2017} but these models have not been compared yet with hot jet and cool surges observed simultaneously.
\\
On the other hand, for another category of MHD models the  important mechanism which drives the jet onset is not the emerging flux itself but the injection of helicity through photospheric motions 
\citep{Pariat2015,Pariat2016}; see further references in the review by \citep{Raouafi2016}.
The presence of shear and/or twist motions  at the base of the closed  non potential region under a preexisting null point induces reconnection with the ambient quasi potential flux  and initiates 
untwisting/helical jets \citep{Pariat2015,Torok2016}.  In some of  these MHD models based on the loss of equilibrium through twisting motions, the thermal plasma parameters of the  jets are not directly considered but suggested  by correspondence  parameters like the plasma $\beta$ \citep{Pariat2016}.  

The observational analysis
from previous studies
 has revealed that the jet evolution could be 
preceded by some wave-like or oscillatory disturbances \citep{Pucci2012, Li2015,
 Bagashvili2018}.
\cite{Pucci2012} analysed the X-ray jets observed by {\it Hinode} 2007 November 02--04 and
found that most of the jets are associated with  oscillations of the coronal emission in bright points (for a recent review of coronal bright points, see \citealt{Madjarska2019}) at the base of the jets.
They concluded that the pre--jet oscillations are  the result of the change  of 
the area or the 
temperature of pre--jet activity region. 
Recently a statistical analysis of pre--jet oscillations of coronal hole jets has been carried out
by \cite{Bagashvili2018}. They reported that 20 out of 23 jets in their study were preceded by 
pre--jet  intensity oscillations some  12--15 mins before  the onset of the jet.  
They tentatively suggested that these quasi periodic intensity
oscillations  may be  the result of MHD wave generation through rapid temperature variations and shear flows associated with  local reconnection events \citep{shergelashvili2006}.

\begin{figure*}
\centering
\includegraphics[width=0.5\textwidth,angle=0]{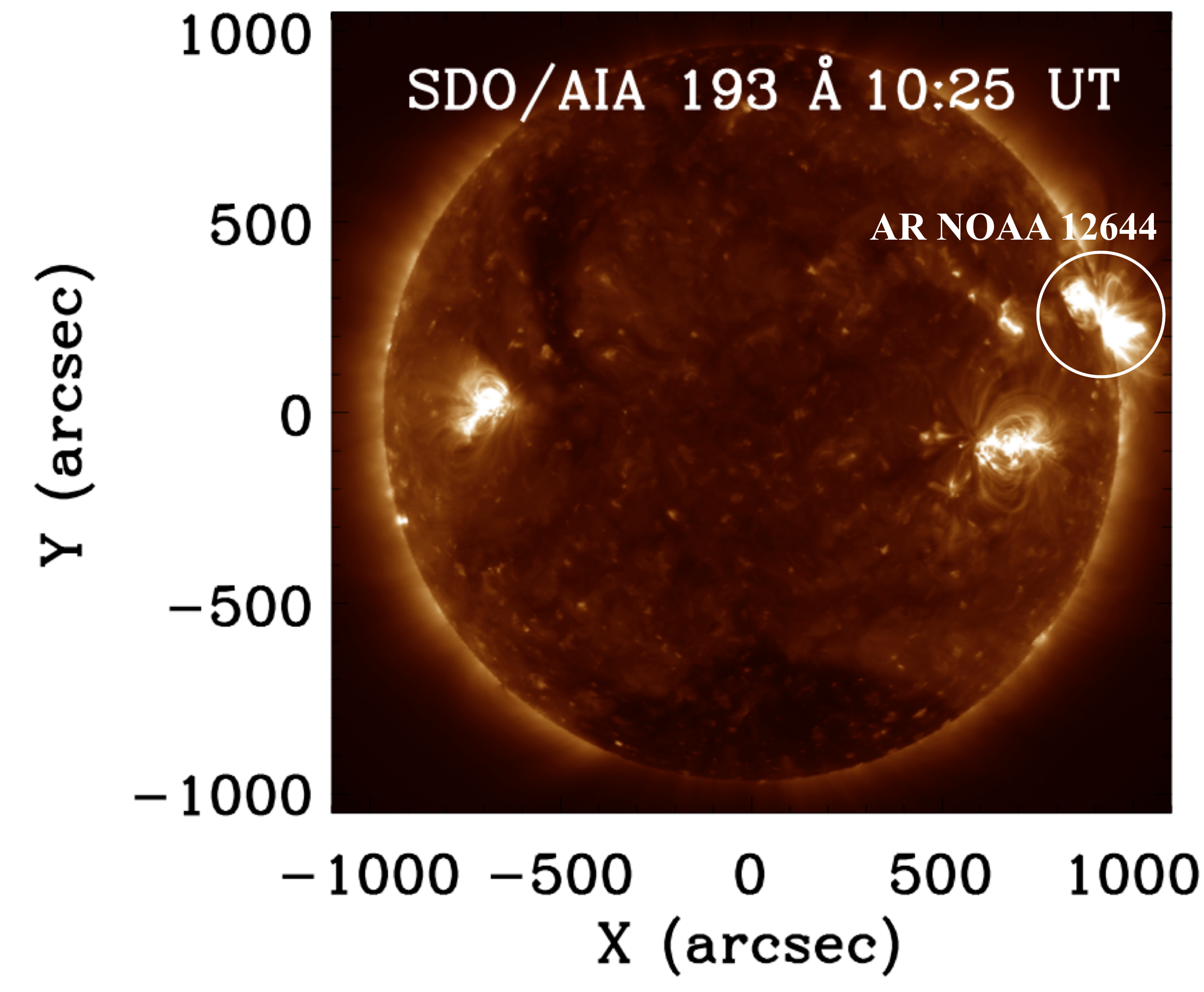}
\caption{Full disk image of the Sun on April 04, 2017.
The solar jets were ejected
from the active region NOAA 12644 shown by the white circle at the west limb.}
\label{full}
\end{figure*}

 Here, we found  a series of 
 jets observed 
 in the hot EUV channels of SDO/AIA as well as 
 in cool temperatures with IRIS slit--jaw images. 
The 
jets were ejected from the 
active region NOAA 12644 on April 04, 2017; on that date, the region was
located at the west limb (N13W91) (Figure \ref{full}).
When passing through the central meridian, this region had shown high jet activity alongside episodes of emerging magnetic flux \citep{Ruan2019}. Its location at the limb in the present observations allows us to visualize the structure of the brightenings from the side and thus facilitates the comparison with the MHD jet models, which motivates the present research. 

 The layout of the paper is as follows. We present the observations and kinematics of jets and identify the reconnected structures  in section \ref{observation}. Pre--jet oscillations are reported in section \ref{ocs}. We discuss 
 our results in section \ref{res}. We conclude that this series of jet and surge observations obtained with a high spatial and temporal resolution match important aspects of the expected behaviour predicted by the
 MHD models of emerging flux. We could identify a candidate location for the current sheet and reconnection site and follow the evolution of the cool surge and hot jets with individual blob ejections. This is a clear 
 case-study for the emerging flux 
 MHD jet  models. 
 
\section{Jets}\label{observation}
\subsection{Observations}
In this study, we select six jet eruptions occurring  in the active region NOAA 12644 
at the western  solar limb on April 04, 2017.
 We use data from  the {\it Atmospheric
Imaging Assembly} (AIA) \citep{Lemen2012} on board  the {\it Solar Dynamics Observatory}
(SDO) \citep{Pesnell2012} and  the {\it Interface Region Imaging Spectrograph} (IRIS) \citep{Pontieu2014}.
AIA observes the full Sun in seven UV/EUV wavelengths 
(94 \AA, 131 \AA, 171 \AA, 193 \AA, 211 \AA, 304
\AA, and 335 \AA\ ) with a pixel size and temporal cadence of $0 \farcs 6$ and 12s,  respectively.
We align the complete data set using the drot\_map routine.
For the bad pixel correction, 
we  process the level 1 AIA data to level 1.5 by 
using the code aia\_prep.pro. These codes are available in {\it SolarSoftWare} (SSW) in IDL platform.
IRIS
provides simultaneously  spectra and images of the 
photosphere, chromosphere, transition region, and
corona which cover a temperature range between 5000 K to 10 MK. 
Slit--jaw Images (SJI) are obtained 
in four different passbands  with a high spatial 
and temporal resolution of $0\farcs16$ pixel$^{-1}$ and
1.5 s respectively. The IRIS data set includes two transition region lines (C II 1330 \AA,  Si IV 1400 \AA), 
one chromospheric line (Mg IIk 2796 \AA), and one photospheric passband 
(in the Mg II wing  around 2830 \AA). 
 We take the  IRIS level 1.5 data from the data archive at 
http://iris.lmsal.com/search. 
The level 1.5 data is corrected for 
dark current and  we remove the FUV background data by iris\_prep.pro in SSWIDL.
The IRIS target was pointed towards the 
active region NOAA 12644 at the western limb with a field of view
of 126$\arcsec$ x 119$\arcsec$ between  11:05:38 UT and  17:58:35 UT. 
For our current study, we  use the SJIs in the C II 
and 
Mg II k bandpasses obtained  with a cadence of 16s.
 The SJIs picture the chromospheric plasma around 10$^4$ K.

\begin{figure*}
\centering
\includegraphics[width=0.8\textwidth]{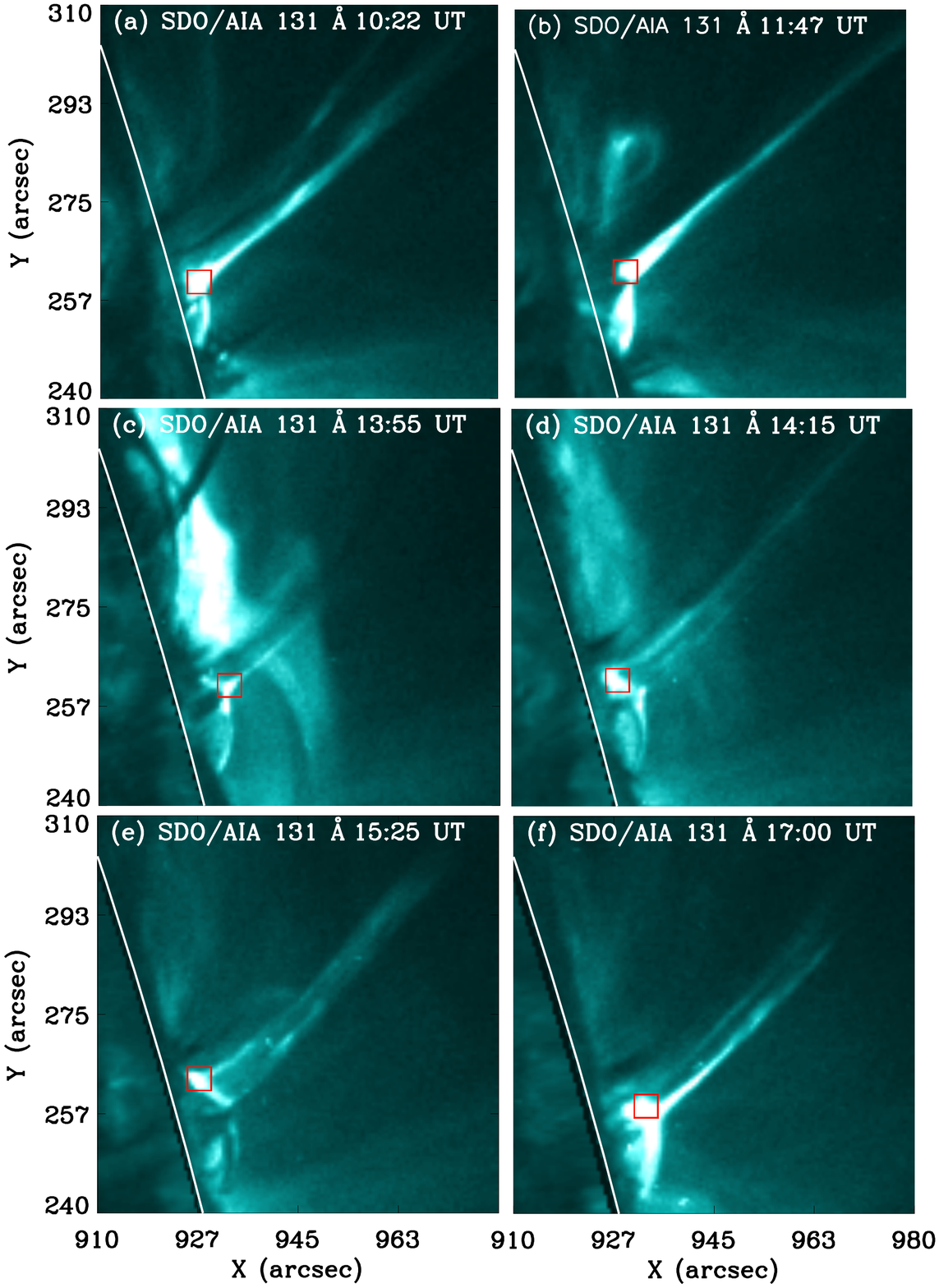}
\caption{Six solar jets ({\it Jet1}--{\it Jet6}) in AIA 131 \AA\ 
filter. The red square in each panel
shows the position at which the pre--jet oscillations are measured. The limb is indicated by the white circle part in each panel.}
\label{evolution}
\end{figure*}

\subsection{Characteristics of the jets}\label{sec:characteristics}
\label{morpho}
On  April 04, 2017, active region jets were 
observed at the limb between 02:30--17:10 UT with AIA. The movies in different wavelengths of AIA  (131 \AA, 171 \AA, and 304 \AA) reveal that there are two sites of plasma ejections (jets) along the  limb. First, there is a northern site ([921$\arcsec$, 264$\arcsec$]), where the jets are straight and have their base located behind the limb and hence concealed by it. Second, there is a site in the south of the field of view ([931$\arcsec$, 255$\arcsec$]) in which the jets have their base over the limb.
Therefore we study in the present paper the six main jets originated in the southern site occurring after 10:00 UT. Five of them were also observed by IRIS, whereas the first of them occurred before the IRIS observations. 
These jets reach an altitude 
between 30 and 70 Mm; their recurrence period is around 80 mins, with the exception of two jets which were separated by only 15 min. In the movies we also see many small jets reaching less than 10 Mm height both before and in between the main jets. 
The jets observed in AIA 131 \AA\ are shown in 
Figure \ref{evolution} (a--f) and in an accompanying animation (MOV1). 
 The first main jet, {\it Jet1}, reaches its peak at $\approx$ 10:22  UT 
 with  an average speed of 210 km s$^{-1}$ (panel (a)). {\it Jet2} (panel b) starts at 11:45 UT and reaches its maximum extent at $\approx$ 11:47 UT. 
 In the movie (MOV1) we note  a large 
 filament eruption located in the northern site of the jets which erupts $\approx$ 13:30 UT and falls back after reaching its maximum height. 
Moreover, we could see that the jet and the filament 
 are not associated with each other. 
{\it Jet3} and {\it Jet4} (panel b and c, respectively) reach  their maximum altitude 
at 13:55 UT and 14:15 UT respectively.
{\it Jet5} (panel e) erupts 
with a broader base and 
 reaches its maximum height at $\approx$ 15:25 UT.
We see a fast  lateral extension
 of the jet base along a bright loop.
{\it Jet6} (panel f) is  ejected at $\approx$ 16:57 UT.  A second instance of filament eruption is observed during the peak phase of {\it Jet6} starting again at the same 
 location of the first one.
 In this case the erupted filament material seems to merge  later
 with the jet material and 
is ejected in the same direction.
However, here the jet is not launched by the filament eruption, because it is not at the jet footpoint. 
\begin{figure*}
\centering
\includegraphics[width=0.7\textwidth,angle=0]{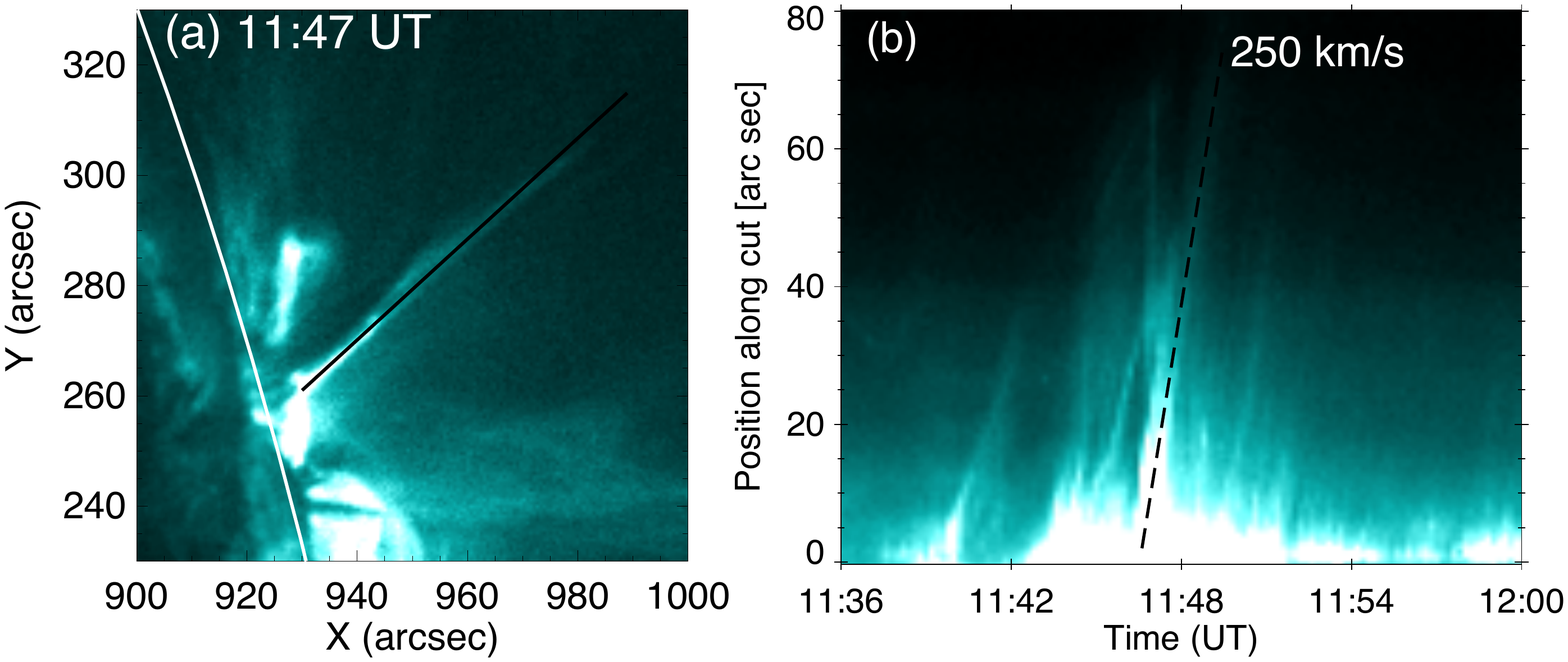}
\vspace*{-2cm}
\caption{An example of timeslice analysis of the jet 1, used
for velocity and height calculations in AIA 131 \AA. In panel (a) the 
solid black line is the slit location, which we use to make the 
height--time plot (b).}
\label{timeslice}
\end{figure*}

\begin{figure*}
\centering
\includegraphics[width=0.70\textwidth,angle=90]{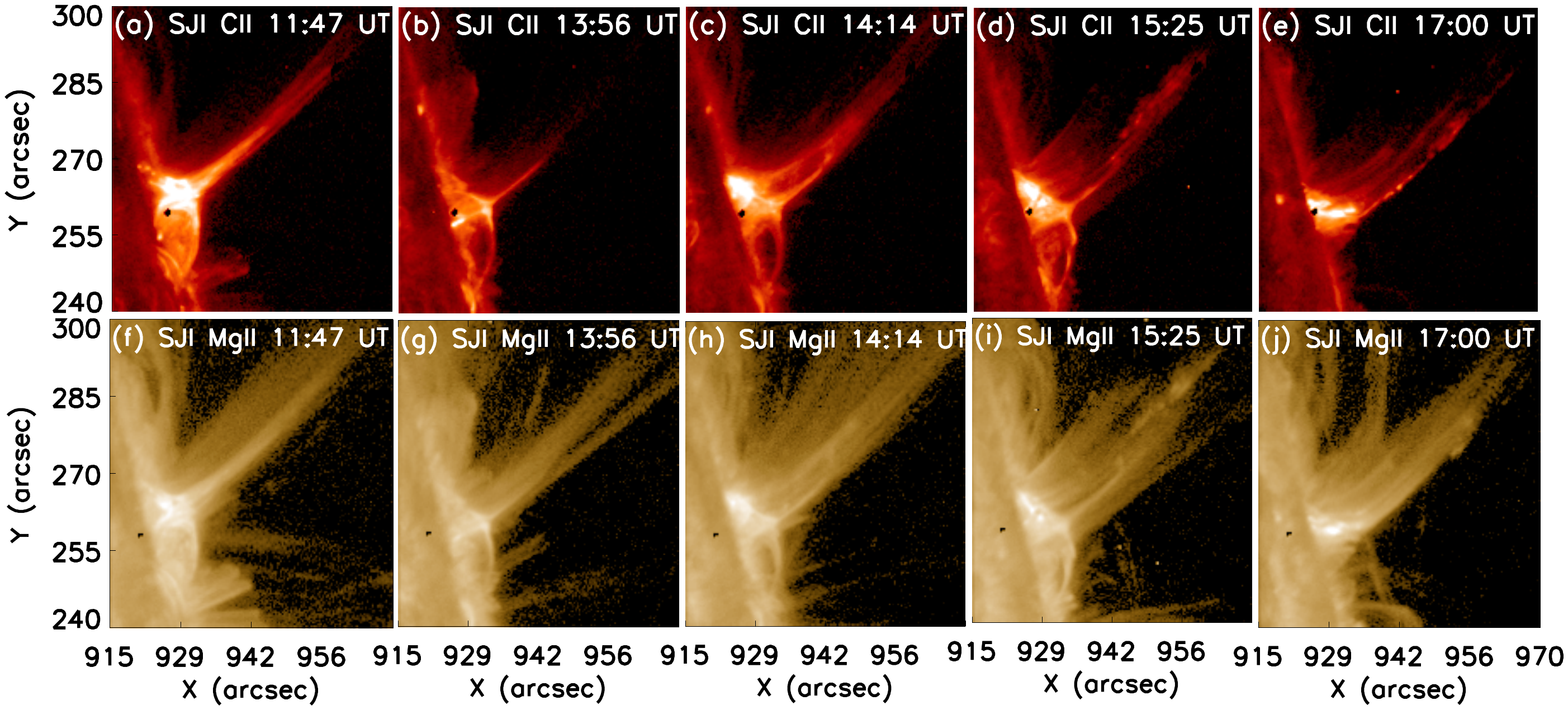}
\vspace*{-1.5cm}
\caption{IRIS observed the active region from 11:05 UT to 17:58 UT.
 It covers five jets in our present analysis in CII (top) and MgII k (bottom)
 lines. 
The black points in the top panels are produced by extreme saturation, which
we used for a better visibility of jets.}
\label{iris}
\end{figure*}

We have computed various physical parameters, namely, height, width, lifetime, speed
 of these jets using the AIA~131~\AA\ data. 
 For the velocity calculation, 
we calibrated
height--time 
of each jet in AIA~131~\AA\ fixing a slit in the middle of the jet plasma flow and calculating the average speed in the flow direction. 
An example of height--time calculation is shown in Figure \ref{timeslice} for {\it Jet2}. 
All computed physical parameters  are listed in Table \ref{table1}.
The maximum height, average speed, width, and lifetime of the observed 
jets vary in the ranges $\approx$ 30--80 Mm, 200--270 km s$^{-1}$, 1--7 Mm, and 
2--10 min respectively.

{\it Jet2}--{\it Jet6} were also observed by IRIS in two wavelength passbands, namely, CII (top row of Figure \ref{iris}) and MgII k (bottom row).
The high spatial resolution of IRIS allowed us to make a clear identification of what looks like a null--point structure at a 
height of $\approx$ 6 Mm. In the CII filter we see bright loops above a bright half dome in the northern site of the jet footpoints. 
In the Mg II filter, the northern part of the dome is also bright.
 We  find jet strands all over the northern side of the dome, like a collection of sheets. We will discuss about these jet strands, which are infact cool jets/surges with a lower velocity in section \ref{theory}.
In AIA 131 \AA\ we see clearly, for all the jets, a  bright area which could correspond to a current sheet (CS), possibly containing a null point, with underlying bright loops shaping a dome (Figure \ref{evolution}). However we notice that the bright dome and loops are located on the southern side of the tentative current sheet, whereas the bright loops  in IRIS C II are rather on its northern side. In the following, for simplicity, when referring to observations of this candidate current sheet and possible null point we will sometimes call them 'the null point' even though there is clearly no way in which one could detect a zero of the magnetic field (nor the intensity of the electric current) in those temperatures with present observational means.
Moreover  in  all the hot channels of AIA (131 \AA\, 193 \AA, 171 \AA, 211  \AA) 
and IRIS  C II and Mg II SJIs 
the jets have an anemone (``Eiffel--Tower" or ``inverted--Y") structure, with a loop at the base and elongated jet arms (see Figures \ref{evolution} and \ref{iris})
as reported in previous events \citep{Nistico2009, Schmieder2013, Liu2016}.


In AIA 131 \AA\ we could also see that between the first and the last jet eruption, the tentative current sheet and the jet spine move towards the south--west direction (Figure \ref{evolution}). More precisely, by following the motion of the point with maximum intensity, we determined a drift of 5 arcsec in less than 6 hours.

\subsection{Temperature and emission measure analysis}\label{DEM}
\begin{figure*}
\centering
\includegraphics[width=0.8\textwidth]{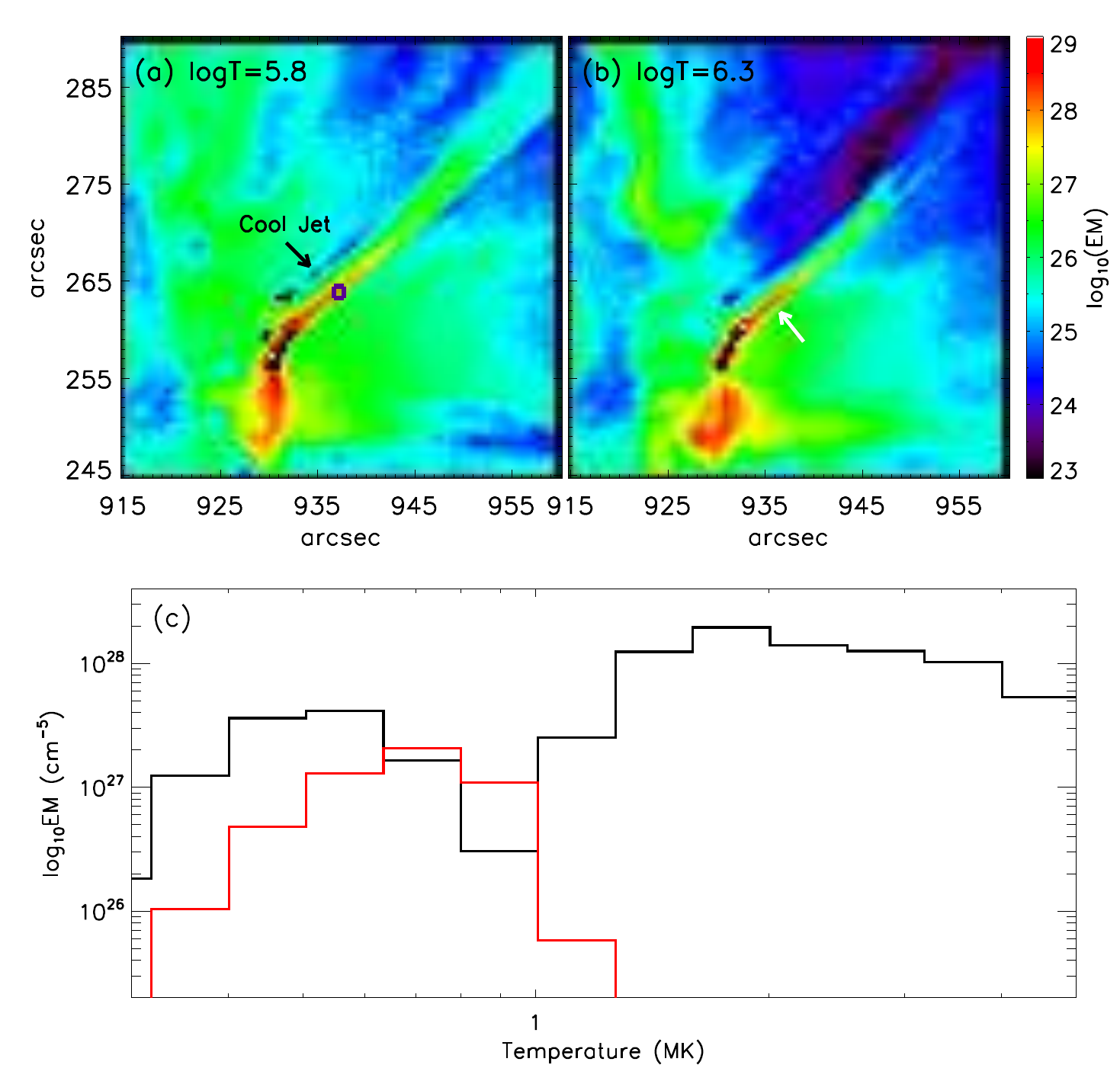}
\caption{Top: Two maps of the active region {\it Jet2} at 11:45 UT,
at two different temperatures (log T= 5.8 and 6.3). The blue square in panel (a) is showing the
location which we  use for the emission analysis at the jet spire. 
The black arrow in panel (a) indicates a weak region of EM that we call the cool jet, the white arrow in panel (b) the hot jet.
 Bottom: In panel (c) the red line shows the temperature at
the location of the  blue box
in panel (a) before the first jet ejection on April 04 2017 and
the black line shows the temperature of solar active region jet at
11:45 UT.}
\label{emission}
\end{figure*}

We have investigated the distribution of the temperature and emission measure (EM)  at the jet spire for all jet events. 
We performed the differential emission measure (DEM)
analysis with the regularized inversion method
introduced by \cite{Hannah2012} using six
AIA channels (94 \AA, 131 \AA,
171 \AA, 193 \AA, 211 \AA, and 335 \AA.
After this process we find the regularized DEM maps as a function of temperature.
We use a temperature range from 
log T(K) = 5.5 to 7 with 15 different bins of width $\Delta$ log T = 0.1. 
 We calculated the EM and lower limit of electron density in the jet spire 
using n$_e$ = $\sqrt{EM/h}$, with $h$ the jet width, assuming that the filling factor equals unity. 
These EM values were obtained by integrating the DEM values over the temperature range
log T(K) = 5.8 to 6.7.
We chose a square box to measure the EM and density 
at the jet spire and at the same location before the jet activity
for each jet.
The example for DEM analysis of 
{\it Jet2} is presented in Figure \ref{emission}, which represents the DEM maps 
at two different temperatures, namely log T (K) = 5.8 (panel a)
and 6.3 (panel b), at 11:45 UT.
We investigate the temperature variation at the jet spire 
during the jet and pre--jet phase. 
 During the  pre--jet phase for {\it Jet2} 
the log EM and the electron density values were 27.3 and 2 x 10$^9$ cm$^{-3}$, whereas for the jet phase the values were 28.1 and 8.6 x 10$^9$ cm$^{-3}$ respectively.
Thus, during the jet evolution the EM value increased by over one order of magnitude and the electron density increased by a factor three at the jet spire. 
We find that the EM and density values increased during the jet phase
in all six jets. The values for all jets are listed in Table \ref{table1}.
\subsection{ Identification of observed structural elements}\label{theory}
\begin{figure*}
\centering
\includegraphics[width=0.7\textwidth,angle=0]{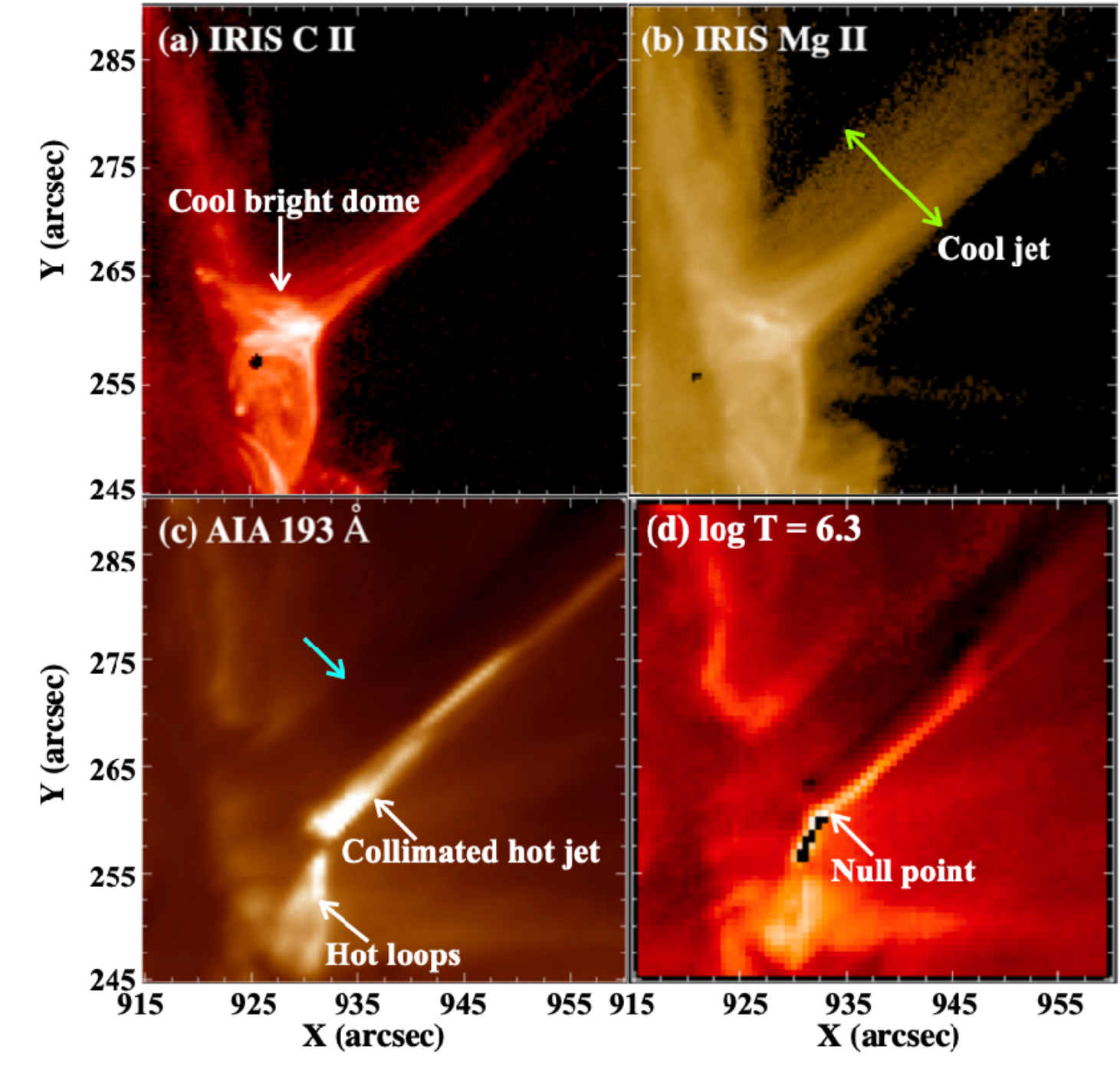}
\caption{Example of {\it Jet2} at 11:45 UT 
observed with IRIS in panel a,b and AIA 193 \AA\ in panel c. Cool bright dome in the northern side of the null point is shown with a white arrow in panel a. We note the broad cool jet in panel b (green arrow), the collimated narrow hot jet with hot loops (white arrows) and absorption area (cyan arrow) in panel c, the null--point and the long bright current sheet (CS) are indicated by the white arrow in panel d. The black points are the saturated areas in panel d.} 
\label{loop}
\end{figure*}

\begin{figure*}
\centering
\vspace*{-2cm}
\includegraphics[width=0.7\textwidth,angle=0]{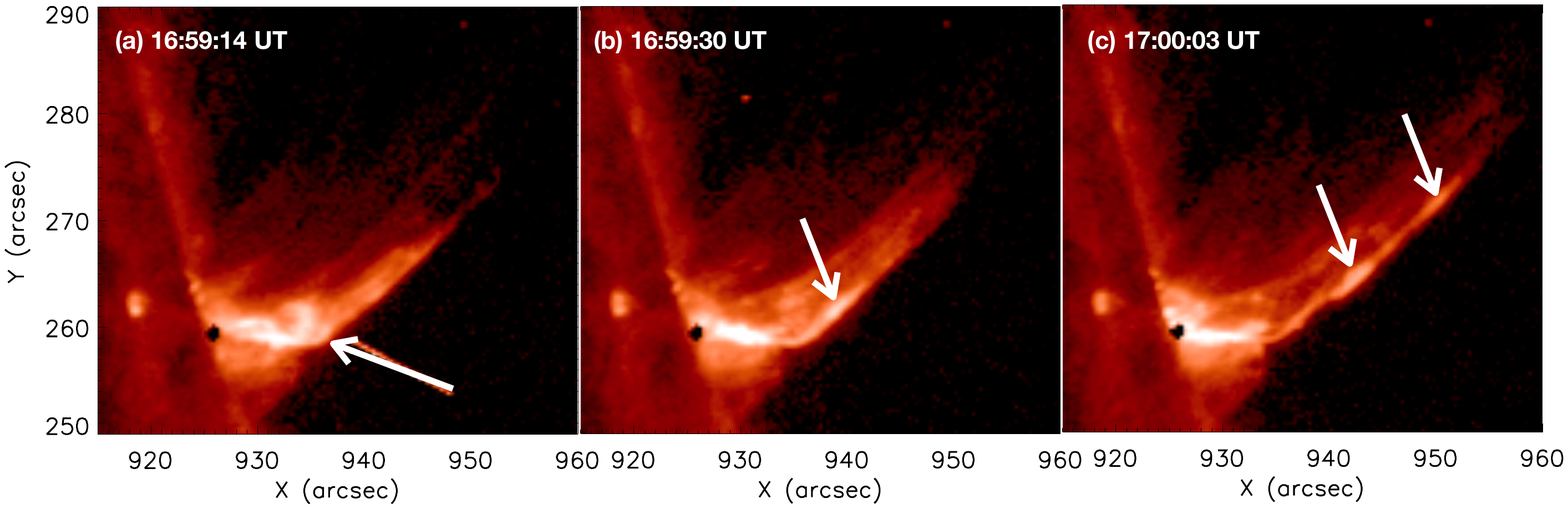}
\vspace*{-2.8cm}
\caption{
Kernels of brightening moving along the {\it Jet6} observed in the IRIS SJI in CII wavelength range (see white arrows). The kernels could correspond to untwisted plasmoids.}
\label{kernel}
\end{figure*}

In Section \ref{morpho} we have discussed the morphology of the jets observed with AIA and IRIS. The region below the jet, as seen in different wavelengths, has a remarkably clear structure, resembling those discussed in theoretical models of the past years. For identification with previous theoretical work, in Figure \ref{loop} several structural elements are indicated for the case of the {\it Jet2} observations.  
 In IRIS CII (Figure \ref{loop}, panel a) the brightenings below the jet delineate a double--chambered vault structure, with the main brightening being located in  the northern part 
of the base of the jet.  Only narrow
loops are seen above the southern part of the vault in this wavelength.
In the other chromospheric line, IRIS Mg II, we see (panel b) roughly the same scenario, although the general picture is rather fuzzier. The jet, in particular, is no longer narrow but formed by parallel strands issuing from the edge of the northern part of the vault, similar to a comb (Figure \ref{loop} panel b). 
%
%
 The assumption of a double-vault structure below the jet is reinforced when checking both the hot-plasma observations (AIA~193~\AA, panel c) and the temperature map obtained through the DEM analysis explained in the previous section (panel d).
%
In those two panels, the southern loops are shown to be bright and hot structures, and the same applies to the point right at the base of the jet, where the temperature reaches $10^6$ K. Additionally, we observe bright kernels moving from time to time along the jets and more clearly visible in {\it Jet4}, {\it Jet5}, and {\it Jet6}. An example of kernels of brightening moving along the {\it Jet6} in IRIS CII is presented in Figure \ref{kernel}. We have computed the velocities of the kernels and find that they are comparable to the mean velocities of the cool jet (45 km s$^{-1}$). The time between  the ejection of two kernels is less than 2 minutes.

  The foregoing structural elements seem to correspond to various prominent features in the numerical 3D models of \citet{Moreno2008} and \citet{Moreno2013}, or in the more recent 2D models of \citet{Nobrega2016, Nobrega2018}, all of which study in detail the consequences in the atmosphere of the emergence of magnetized plasma from below the photosphere. One can identify the bright and hot plasma apparent in the observations at the base of the jet with the null point and CS structures resulting in those simulations (see the scheme in Figure~\ref{null}, right panel): the collision of the emerging magnetized plasma with the preexisting coronal magnetic system leads, when the mutual orientation of the magnetic field is sufficiently different, to the formation of an elongated CS harboring a null point and to reconnection.
As a next step in the pattern identification, the hot plasma loops apparent in the southern vault in the AIA~193~\AA~image and the temperature panels of Figure~\ref{loop} should correspond to the hot post-reconnection loop system in the numerical models (as apparent in Figures~3 and 4 of the paper by \citealt{Moreno2008}, or along the paper by \citealt{Moreno2013}). 
On the other hand, the northern vault appears dark in AIA~193~\AA, and has lower temperatures in the DEM analysis. This region could then correspond to the emerged plasma vault underlying the CS in the numerical models: the magnetized plasma in that region is gradually brought toward the CS where the magnetic field is reconnected with the coronal field.

\begin{figure*}
\centering
\vspace*{-2.0cm}
\includegraphics[width=0.7\textwidth,angle=0]{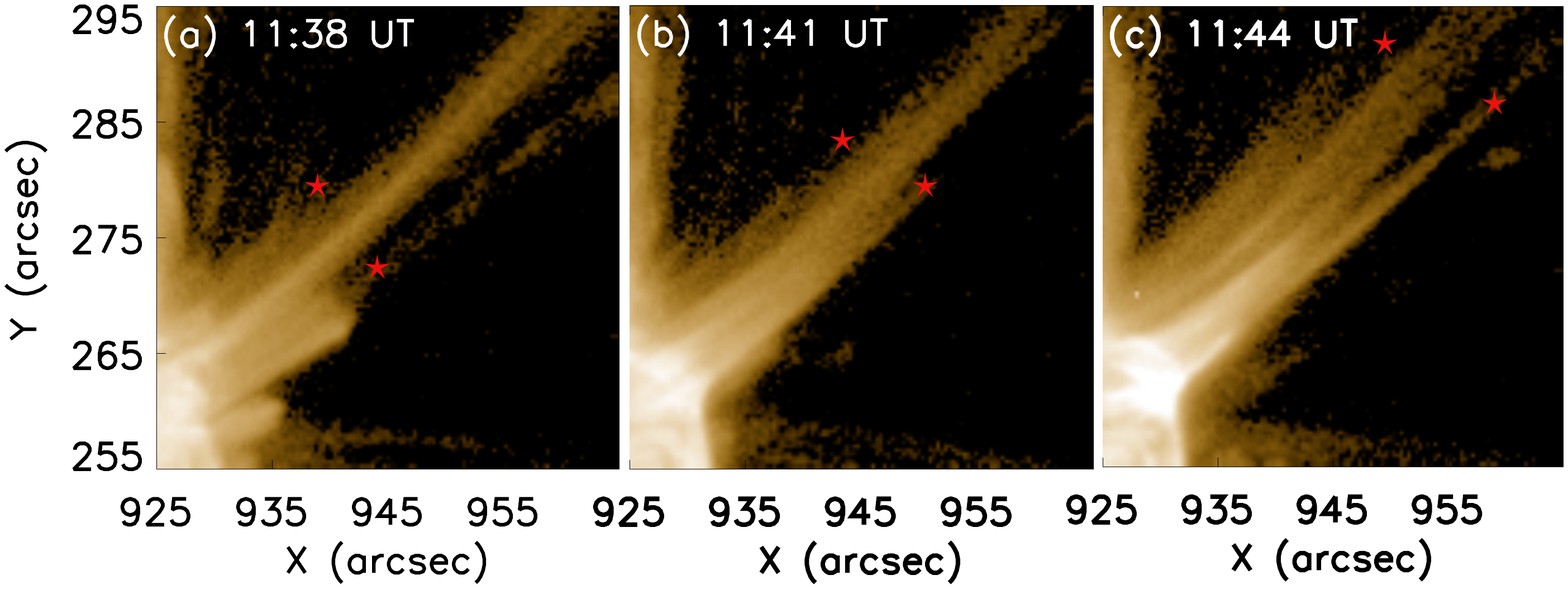}
\vspace*{-2.5cm}
\caption{
The evolution of cool plasma material along both sides of the hot jet ({\it Jet2}) in IRIS MgII wavelength. The red star shows the leading edge of the cool material ejecting with an average speed of 45 km s$^{-1}$.}
\label{cool}
\end{figure*}

Additional features in the observation that fit in the foregoing identification are the following:
\\
(a) As time proceeds the northern chamber decreases in size while the southern chamber grows.
In our observations in the beginning phase of the jets (for instance; {\it jet2} at 11:30 UT) the area of the northern and southern vaults is 1.4 x 10$^{18}$ and  1.16 x 10$^{18}$ cm$^{2}$, respectively, and during the jet phase (11:47 UT), they become
 $1.05 \times 10^{18}$ and $2.2 \times 10^{18}$ cm$^{2}$, respectively.
 This suggests that while the reconnection is occurring, the emerging volume is decreasing whereas the reconnected loop domain grows in size, as in the emerging flux models \citep{Moreno2008, Moreno2013, Nobrega2016}.
\\
(b) A major item for the identification of the observation with the flux emergence models is the possibility 
that we also observe a wide, cool and dense plasma surge ejected in the neighborhood of the vault and jet complex (see movie in C II attached as MOV2).
This  wide laminar jet is observedin the Mg~II  IRIS filter  as an absorption sheet parallel to the hot jet in AIA 193 \AA. The evolution of the cool material along both sides of the hot jet in the IRIS Mg II channel is presented in Figure~\ref{cool} and the leading edge of the cool part is indicated by red stars. 
The cool ejection is generally less collimated than the hot jet and is seen to first rise and then fall, similarly to classical H$\alpha$ surges.
The velocities measured along the cool sheet of plasma in Mg II are $\approx$ 45 km s$^{-1}$.  
 The ejection of cool material next to the hot jets is a robust feature in different flux emergence models \citep{Yokoyama1996, Moreno2008, Nishizuka2008, Moreno2013, MacTaggart2015, Nobrega2016, Nobrega2017, Nobrega2018}. The cool plasma in the models is constituted by matter that has gone over from the emerged plasma domain to the system of reconnected open coronal field lines without passing near the reconnection site, that is, just by flowing, because of flux freezing, alongside the magnetic lines that are being reconnected at a higher level in the corona. All those models report velocities which match very well the observed value quoted above.
 \\
(c) The observed kernels in Figure \ref{kernel} could be plasmoids  created in the CS during the reconnection process. In some of the flux emergence models just discussed, plasmoids are created in the CS domain (see, for example \citealt{Moreno2013}), and they are hurled out of the sheet probably via the melon-seed instability \citep{Nobrega2016}, even though they are not seen to reach the jet region. On the other hand,  in the 2D jet model by \citet{Ni2017}, plasmoids are created in the reconnection site that maintain their identity when rising along the jet spire, possibly because of the higher resolution afforded by the Advanced Mesh Refinement used in the model; this is in agreement with the behavior noted in the present observations as well as in the previous observations of \citet{Zhang_Ji_2014} and \citet{Zhang2016} mentioned in the introduction. Plasmoids are also generated in the model by \citet{Wyper2016}, which is 
a result of footpoint driving of the coronal field rather than flux emergence from the interior.
%
%
On the other hand, the formation of the kernels could follow the development of the Kelvin–Helmholtz instability (KHI). The KHI can be produced when two neighboring fluids  flow in same direction with different speed \citep{Chandrasekhar1961}. This instability may develop following the shear between the jet and its surroundings. For details about this sort of process in jets and CMEs see the review by \cite{Zhelyazkov2019}.
\\
(d) The main brightening at the top of the two vaults seems to be changing position systematically in the observations.
There is a shift in the south--west direction as time advances, and the same displacement is apparent in AIA 131 \AA\ (see Figures \ref{evolution} and \ref{iris}), possibly marking the motion of the reconnection site. Such type of observations are also reported in the study of \cite{Filippov2009}. This shift may be used to compare with the drift of the null point position detected in the MHD models.

(e) We also notice a significant rise of the brighter point (null point)
between different jet events. The rise of the reconnection site as the jet evolution advances has been found in the MHD emerging flux models of \citet{Yokoyama1995, Torok2009}.
In the present case, it may be because during each jet event the reconnection process causes a displacement of the null point and jet spine. In this way the next jet event occurs in a displaced location as compared with the previous jet. This could indicate that the magnetic field configuration has some reminiscences of the earlier reconnection and behaving in the same manner afterwards. Another possible reason for this shifting could be as a result of the interaction between different quasi–separatrix layers (QSLs) as suggested by \citealt{Joshi2017}. However, in the present case because of the limb location of the active region, we could not compute the QSL locations.

\section{Oscillations before  the jet activity}\label{ocs}


In \S\ref{sec:characteristics} we mentioned that before and in between the six main jets we also observed many small jet-like ejections, with length less than $10$ Mm. Also, in the AIA  131~\AA\ movie we clearly see many episodic brightenings related to the small jets. In the present section we would like to investigate different properties, like the periodicity, of these features. To that end,  
we select a square of size $4 \times 4$ arcsec at the base of the jets where the intensity is maximum, in  the AIA  131 \AA\ data, as shown in  Figure \ref{evolution} and calculate the mean intensity inside the square in the AIA 131 \AA\ channel. 
We compute the relative intensity variation in the base, after normalization by the quiet region intensity. 
 We find that the oscillations start at the jet base some 5--40  min before the  main jet activity.

\begin{figure*}
\centering
\includegraphics[width=0.7\textwidth,angle=0]{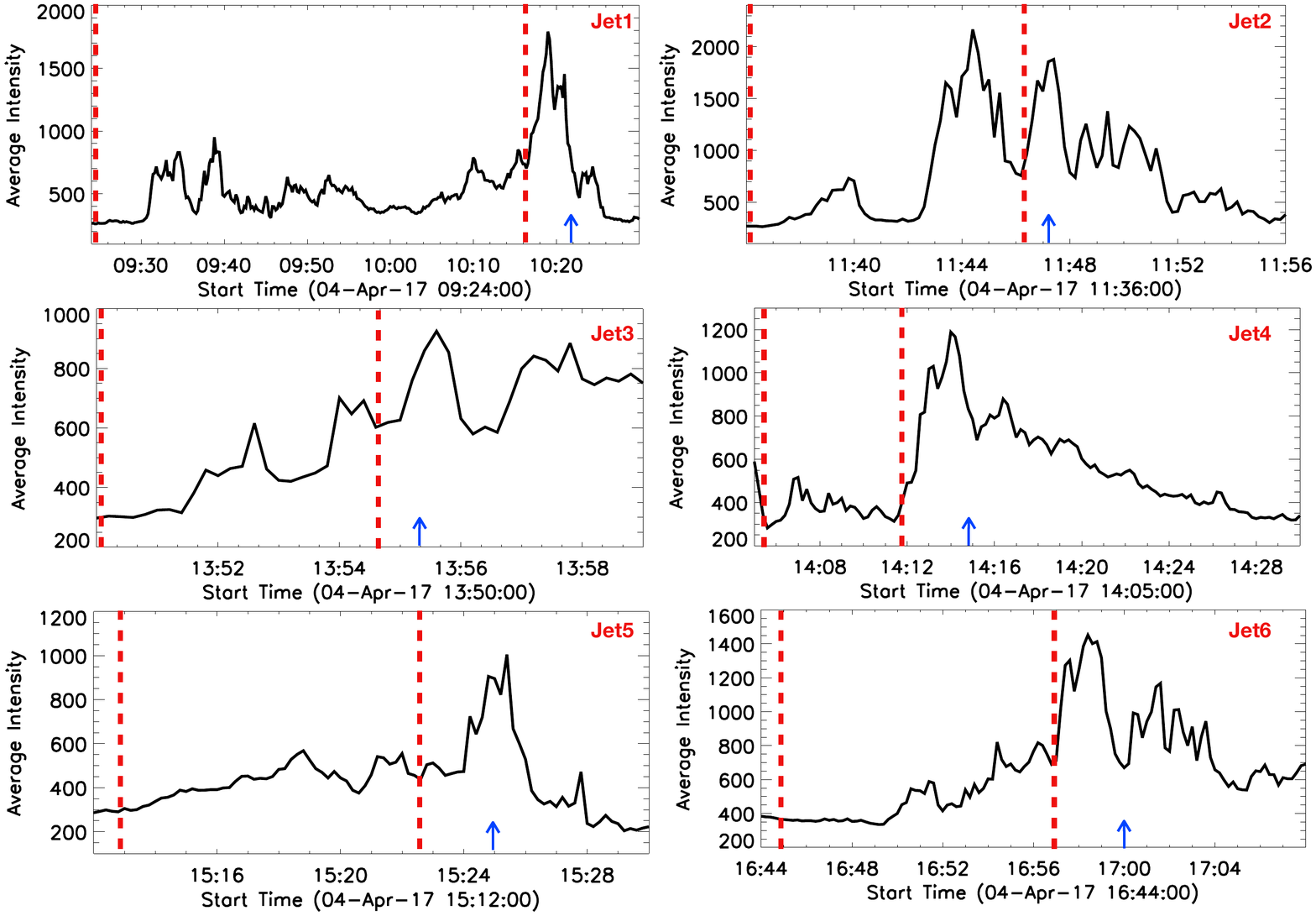}
\caption{Intensity distribution during pre--phase of recurrent jets at the base of each jet in AIA 131 \AA.
The location of the jet base of each jet 
is displayed in Figure \ref{evolution} (see red squares).
In each panel, a blue arrow indicates the peak phase of the main jet. The small intensity peaks before  each main jet are related to small jet ejections (10
Mm height) coming from the same location. The two vertical red dashed
lines in each panel show the duration of the pre--jet intensity oscillation that is used for 
the wavelet analysis.
}
\label{intensity}
\end{figure*}

Figure \ref{intensity} shows the intensity distribution at the jet base for all the jets before and during the jet eruption and the pre--jet phase is shown in between two vertical red dashed lines. The right red dashed lines indicate the starting time of the main jets. The blue arrows indicate the time of the maximum elongation of the main jets. We note that the maximum of the brightening at  the jet base  does not  always  coincide exactly  with the start of the jet neither with the maximum extension time.  
In most of the cases the maximum brightening  occurs  before the 
peak time of the jets by a few minutes.
For the smaller jets it is nearly impossible to compute the delay between  brightenings and jets. They appear to be in phase with the accuracy of the measurements.

To calculate the time period of these pre--jet oscillations, we apply a  wavelet analysis technique. For the significance of time periods in the wavelet spectra, we take a significance test 
into 
account and the levels higher than or equal to 
95\% are labeled as real. The significance test and the 
wavelet analysis technique is well described by \cite{Torrence1998}.
The cone of influence (COI) regions
make 
an important background for the edge effect at the 
start and end point of  the time range \citep{Tian2008, Luna2017}.  

\begin{table}[h!]
  \caption{Physical parameters of six studied hot jets.}
\bigskip
\label{table1}
\setlength{\tabcolsep}{0.7pt}
\begin{tabular}{c|c|c|c|c|c|c|c} 
      \hline
      \text{Jet} & \text{Jet start}& \text{Jet peak} & \text{Max} 
& \text{Average} &\text{T} &\text{EM}&
  \text{Oscillation}\\
      \text{no.} & \text{time} & \text{time} & \text{height} 
& \text{speed} &\text{ } &\text{(10$^{28}$}&
  \text{period}\\
  \text{ } & \text{(UT)} & \text{(UT)} & \text{(Mm)} 
& \text{(km s$^{-1}$)} &\text{(MK)} &\text{ cm$^{-5})$}&
  \text{(min)}\\
      \hline
       
       1 & 10:15 & 10:22 & 80 & 210 & 1.4 & 1.4 & 6.0\\
       2 & 11:46 & 11:47 & 50 & 245 & 1.8 & 1.9 & 1.5\\
       3 & 13:54 & 13:55 & 40 & 265 & 1.4 & 1.5 & 2.5\\
       4 & 14:12 & 14:15 & 50 & 250 & 1.8 & 1.1 & 2.0\\
       5 & 15:23 & 15:25 & 55 & 235 & 1.8 & 1.3 & 4.0\\
       6 & 16:57 & 17:00 & 70 & 220 & 1.8 & 2.0 & 2.5\\
      \hline
    \end{tabular}
\end{table}

\begin{figure*}
\centering
\includegraphics[width=0.4\textwidth,angle=90]{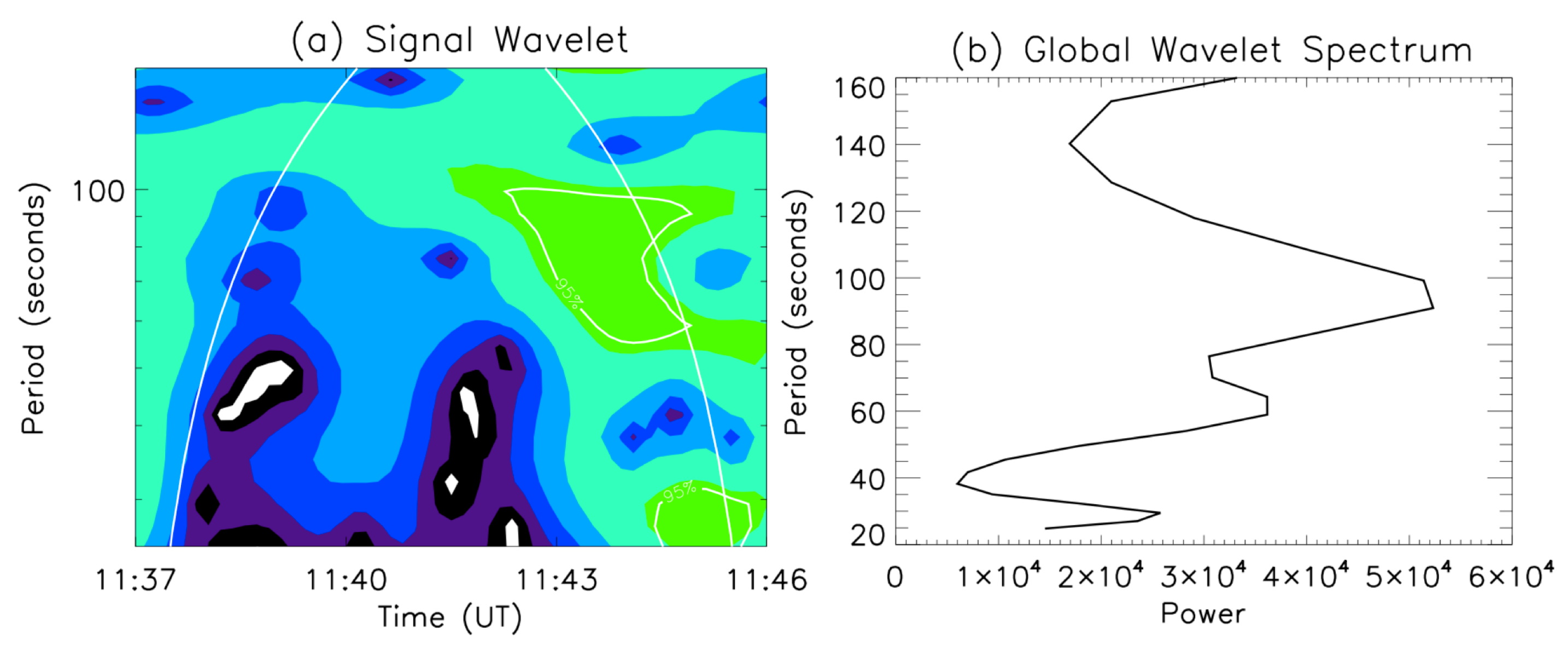}
\caption{Panel (a): An example of wavelet spectrum for the pre–jet intensity oscillations
for {\it Jet2}. The solid thick white contours (around the green surface) are the regions with the value of wavelet function larger than the 95\% of its maximum value. 
The area which is outside the parabolic COI is the region where the wavelet analysis is not valid.
Panel (b): Global wavelet spectra for the distribution of
power over time. The highest peak is corresponding to the time period of the pre--jet intensity oscillations, {\it i.e.} 1.5 min for {\it Jet2}.
}
\label{wavelet}
\end{figure*}

\begin{figure*}
\centering
\includegraphics[width=0.65\textwidth,angle=0]{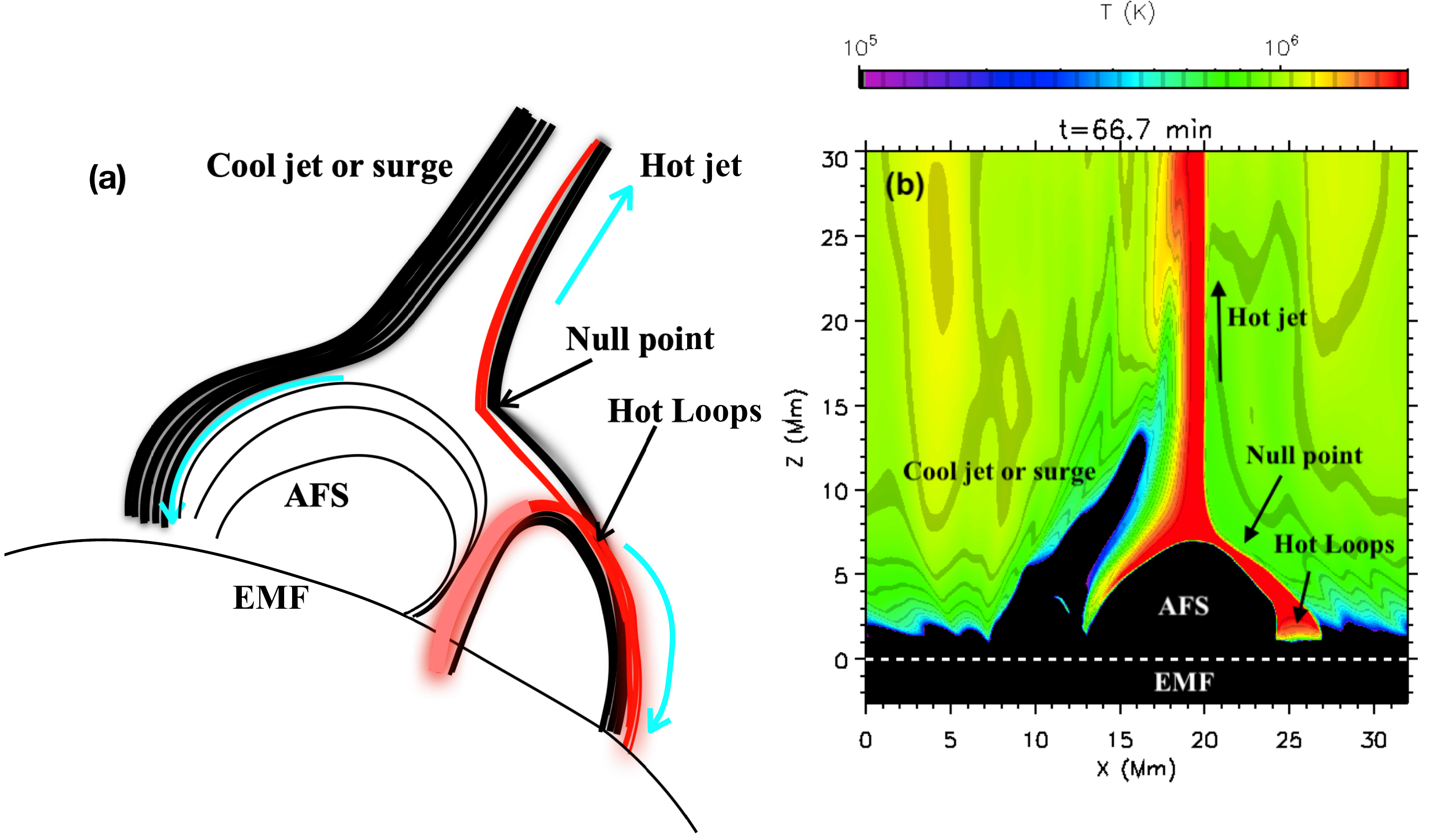}
\caption{
Panel (a): Schematic view of the 3D jet derived from \cite{Moreno2008},
showing the location of the null point, the cool surge and the hot loops
next to the AFS.
The cyan arrows indicate the direction of the flows.
The red lines indicate the presence of hot plasma. The black lines are magnetic field lines.
Panel (b): 
Temperature map of one of the numerical experiments by \citet{Nobrega2017, Nobrega2018} showing the hot jet and the cool surge
(an animation of this panel is available online).
In both panels, the region of the convection zone where the new magnetic flux has emerged (EMF)
is also indicated.}
\label{null}
\end{figure*}

The wavelet analysis of the intensity fluctuation at the jet base 
shows that the oscillation period for these pre--jet intensity varies between 1.5 min and 6 min; the current values obtained are presented in the last column of Table \ref{table1}. 
An example of wavelet spectrum for the pre--jet activity for {\it Jet2} is presented in Figure \ref{wavelet} (a). The COI region is the outer 
area of the white parabolic curve.
The global wavelet spectrum in panel (b) shows the distribution of power spectra over time.  
 \cite{Bagashvili2018} investigated the intensity at the base of several jets issued in a coronal hole and  obtained similar results concerning the  periodicity and  duration of the oscillations.

\section{Discussion and conclusion}\label{res}
This paper presents observations concerning the structure,
kinematics, and pre--jet intensity oscillations of
 six major jets that occurred on April 04, 2017 in active
   region NOAA 12644. The discussion is based on 
the observational data from AIA and IRIS. 
 A brief summary of our main results is as follows:
 
 \begin{enumerate}
 \item All the jets show  pre--jet intensity oscillations at their
base accompanied by smaller jets. The period of the oscillation
  ranges from 1.5 to 6 min. 

\item The jets are issued from a canopy-like structure with two vaults delineated by the brightenings seen in the different wavelengths. One of the vaults harbors hot loops as seen in the EUV AIA filters and also  in IRIS C II wavelength. The hot jets are accompanied by laminar cool surge--like jets visible in IRIS Mg II and  C II wavelengths.

\item The spatial and temporal pattern of brightenings in the various
  wavelengths show clear similarities with the two- and
  three-dimensional numerical models of \citet{Moreno2008, Moreno2013} and \citet{Nobrega2016}. 
The high brightening overlying the two vaults in the observations, in particular, is suggestive of the null point and CS complex in the models; the two vaults would then correspond to the domains occupied by the emerging plasma and the reconnected hot loops, respectively, in the models.
\item
The cool surge-like jets visible in the IRIS images and in absorption in AIA filters may be the counterpart to the cool ejections that naturally accompany the flux emergence models. Further observed features that are present in flux emergence models are: the ejection of bright kernels from the region identified as the reconnection site, and the shift in the reconnection site towards the south--west direction.



\end{enumerate}

In the following, a discussion of those results is provided:
\\
 
A first significant finding of this study is the observation of pre--jet activity,
in particular in the form of oscillatory behavior. Earlier authors had
  studied the pre--jet activity of quiet region jets observed in the hot AIA filters 
\citep{Bagashvili2018}.  The jets studied by those authors
had their origin in
coronal bright points 
and the bright points
showed oscillatory behavior before the onset of jet
activity.  They reported periods for the pre--jet oscillations of around 3 mins.  Our study deals with active region jets, instead,  also observed in the hot filters of AIA and we find  an oscillatory behavior of the intensity  in a time interval of 5--40 min  prior to the
  onset of the jet. 
  The period of the  intensity oscillation is
  in the range 
  1.5--6 min. These values are consistent with
the results reported by \cite{Bagashvili2018}. They are also
close to  typical periods of acoustic waves in the magnetized solar
atmosphere. This indicates that 
acoustic waves may be responsible
for these observed periods in the occurrence of jets
\citep{Nakarikov2005}. 
Quasi-oscillatory variations of intensity can also be the signature of MHD wave excitation processes  which are generated by very rapid dynamical changes of velocity, temperature and other parameters manifesting the apparent non-equilibrium state of the medium where the oscillations are sustained \citep{Shergelashvili2005, Shergelashvili2007, Zaqarashvili2002}.
 In 3D reconnection regions like the 
 quasi-separatrix layers, a sharp velocity gradient is likely to be present. The impulsiveness of the jets could lead to such MHD wave excitation. The fast change of the dynamics and thermal parameters  at these reconnection sites  should be checked when possible to  prove   the interpretation of the intensity oscillation  by  MHD waves.
  The observed brightness fluctuations could also be due to the oscillatory character of the reconnection processes that lead to the launching of the small jets. Oscillatory reconnection has been found in  theoretical contexts in two dimensions \citep{Craig1991, McLaughlin2009, Murray2009}. The latter authors, in particular, studied the emergence of a magnetic flux rope into the solar atmosphere endowed with a vertical
magnetic field. As the process advances, reconnection occurs in the form of bursts with reversals of the sense of reconnection, whereby the inflow and outflow magnetic fields of one burst become the outflow
and inflow fields, respectively, in the following one. 
The period of the oscillation covers a large range, 1.5 min to 32 min. They concluded that the characteristics of oscillatory reconnection and MHD modes are quite  similar. However, that model is two-dimensional and it is not clear if the oscillatory nature of the reconnection can also be found in general 3D environments.
A second significant point in our study is the comparison of 
the observations of the structures and time evolution of the jet complex with numerical experiments of the launching of jets following flux emergence episodes from the solar interior.
Structures like the  double-vault dome with a  bright point at the top where the jets  are initiated as seen in the hot AIA channels and also in the high-resolution IRIS images mimic the structures found in the numerical simulations of   \citet{Moreno2008} and \citet{Moreno2013}, who solved the MHD equations in
three dimensions to study the launching of coronal jets following the emergence of magnetic flux  from the solar interior into the atmosphere; they also have similarities with the more recent experiments, in two dimensions, of \citet{Nobrega2016}, obtained with the radiation-MHD Bifrost
code \citep{Gudiksen:2011}. In the 3D models, the jet is launched along open coronal field lines
that result from the reconnection of the emerged field with
the preexisting ambient coronal field. Underneath the jet, two vault
structures are formed, one containing the emerging cool plasma and the other a set of hot, closed coronal loops resulting from the reconnection. Overlying the two vaults one finds a flattened CS of Syrovatskii type, which
contains hot plasma and where the reconnection is occurring. The field in the
sheet has a complex structure with a variety of null points; in fact, in its
interior, plasmoids, with the shape of tightly wound solenoids, are seen to
be formed. The reconnection is of the 3D type, in broad terms of the kind
described in the paper by \citet{Archontis2005}. A vertical cut of the 3D
structure, as in Figure 4 of the paper by \citet{Moreno2008}, clearly shows
the two vaults with the overlying CS containing the reconnection
site and with the jet issuing upwards from it. The figures in that paper
contained values for the variables as obtained solving the physical
equations; a scheme of the general structure is provided here as well
(Figure~\ref{null}, left panel). As the reconnection process advances, the
hot-loop vault grows in size whereas the emerged-plasma region decreases,
very much as observed in the present paper.

An interesting feature in the observations is the tentative detection
  of a surge-like episode next to the jet apparent in the IRIS
  Mg-II time series 
    in a region that appears dark, 
   in absorption, in the AIA~193~\AA\ observations.
This ejection of dense and cool plasma next to the hot 
  jet, with the cool matter rising and falling, like in an H$\alpha$ surge,
  also occurs naturally both in the 3D and 2D numerical models cited above
  (and was already introduced by \citealt{Yokoyama1995} in an early 2D
  model). The phenomenon has been studied in depth by \citet{Nobrega2016,Nobrega2017,Nobrega2018}
  using the realistic material properties and radiative transfer provided by
  the Bifrost code, which, in particular, facilitate the study of plasma at
  cool chromospheric temperatures. A snapshot of one of the experiments by
  those authors showing a temperature map and with indication of some major
  features is given in Figure~\ref{null} (right panel). In their model, 
  the magnetic field can accelerate the plasma with accelerations up to $100$ times the solar gravity for very brief periods of time after going through the reconnection site
   because of the high field line curvature and associated large Lorentz force. In the advanced phase of the surge, instead, the cool plasma basically falls with free-fall speed, just driven by gravity, as had been tentatively concluded in observations (see \citealt{nelson2013}). The velocities obtained from the observations in the present paper broadly agree with those obtained in the numerical models.


We conclude that our observations of the six  EUV jets and surges  constitute a clear case-study for comparison with the experiments  developed  to study flux emergence events such as the MHD  models  of \citet{Moreno2008,Moreno2013,Nobrega2016}.
Many observed structures were identified in their models: the reconnection site with two vaults,   hot jets accompanied by surges, ejections of plasmoids in parallel with the development of the cool surges;  the velocity of the hot jets (250 km $^{-1}$) and of the cool surge (45 km s$^{-1}$), in particular, fit quite well with the predicted velocity in the models.

  The similarities between the observations and the numerical models based on magnetic flux emergence are
  no proof, of course, that the observed jets are directly caused by 
  episodes of magnetic flux emergence through the photosphere into the corona: given
  the limb location of the current observations, there is no possibility of
  ascertaining whether magnetic bipoles are really emerging at the
  photosphere and causing the jet activity. However, a jet from this active region that occurred
on March 30, 2017,  was also studied by
\citet{Ruan2019} ; on that date, the active region was at the 
solar disk and the photospheric magnetic field measurements were
reliable. 
Those authors reported that  flux emergence episodes
  were continually occurring in the active region  at that time.
  Although there can be no 
direct proof  through magnetograms, it is likely that flux emergence
  continued to take place as the active region remain being strong and complex until April 04, 2017. Important jets were  also observed  the day before when the region was close to  the limb with  AIA and in H$\alpha$ but we have no IRIS data to observe the fine structures and the null point.

In the future it will be interesting to observe such events with double vault in multi-wavelengths with AIA and IRIS but on the disk to be able to detect magnetic flux emergence  with  HMI magnetograms. 
It will also be very interesting to have the spectra of IRIS just on the  reconnection  site. 
We would like to detect with high accuracy the formation of plasmoids  in the current sheet using the spectral capabilities of IRIS. Such kind of observations can serve to validate the numerical experiments of the theoretical
scientists.

\begin{acknowledgements}
We acknowledge the anonymous referee for the valuable/constructive comments and suggestions. We thank the SDO/AIA, SDO/HMI, and IRIS science teams for granting free access to the data. We thank to Pascal D\'emoulin for important discussions and suggestions.
RJ acknowledges CEFIPRA for the Raman Charpak fellowship under which this work is carried out at Observatorie de Paris, Meudon, France. RJ also thanks to the Department of Science and Technology, New Delhi India, for the INSPIRE fellowship. BS wants to thank Ramesh Chandra for his invitation in Nainital in October 2019. RGA and BS acknowledge for the support of the National French program (PNST) project "IDEES". RC and PD acknowledges the support from SERB-DST project no. SERB/F/7455/ 2017-17. FMI is grateful to the Spanish Ministry of Science, Innovation and Universities for their financial support through project PGC2018-095832-B-I00. DNS thankfully acknowledges support from the Research Council of Norway through its Centres of Excellence scheme, project number 262622. FMI and DNS also gratefully acknowledge support from the European Research Council through the Synergy Grant number 810218 (ERC-2018-SyG).
We thank to Manuel Luna for helping us in wavelet analysis.
\end{acknowledgements}

\bibliographystyle{aa}
\bibliography{references-3}

\end{document}